\documentclass{llncs}

\usepackage{times}
\usepackage{algorithm2e}
\usepackage{graphicx}
\usepackage{gastex}
\usepackage{wrapfig}
\usepackage{paralist}
\usepackage{multirow}
\usepackage{paralist}
\usepackage[T1]{fontenc}
\usepackage{amssymb}
\usepackage{amsmath}
\usepackage{url}
\usepackage{color}

\usepackage{listings}
\lstdefinelanguage{program}{%
  keywords={%
    atomic,%
    assume,assert,call,return,new,%
    false,true,duplicate,restart,lock,unlock,%
    locate,insert,delete,contains,removeRight,rotateRightLeft,
    rcu_read_lock,rcu_read_unlock,synchronize_rcu%
  },
  morecomment=[l]{//},
  morecomment=[s]{/*}{*/},
  morecomment=[n]{(**}{**)},
  mathescape=true,
  escapeinside=`',
}

\usepackage{hyperref}
\usepackage[capitalise]{cleveref}

\usepackage{listings}
\lstdefinestyle{CStyle}{
    language=C,
    showtabs=false,
    tabsize=2,
    basicstyle=\footnotesize
}
\usepackage{verbatim}
\usepackage{nicefrac}

\newcommand{\Prog}{\textit{Prog}}

\newcommand{\TS}{\textit{TS}}
\newcommand{\Init}{\textit{Init}}

\newcommand{\Prop}{P}
\newcommand{\Vars}{V}
\newcommand{\TR}{\textit{Tr}}
\newcommand{\Inv}{\textit{I}}
\newcommand{\pc}{\texttt{pc}}

\newcommand{\pcend}{{\textit{end}}}
\newcommand{\pcloopone}{{\textit{loop}_1}}
\newcommand{\pclooptwo}{{\textit{loop}_2}}
\newcommand{\pcstart}{{\textit{start}}}
\newcommand{\Progcomp}{P_{M \| \Prog}}
\newcommand{\TScomp}{\TS_{M \| \Prog}}

\Crefname{algorithm}{Alg.}{Alg.}

\author{Sharon Shoham\thanks{This publication is part of a project that has received funding
from the European Research Council (ERC) under the European Union's Horizon 2020
research and innovation programme (grant agreement No [759102-SVIS]).}}

\title{Undecidability of Inferring Linear Integer Invariants}

\institute{Tel Aviv University}

\begin{document}
\pagestyle{plain}
\maketitle

\begin{abstract}
We show that the problem of determining the existence of an inductive invariant in the language of quantifier free linear integer arithmetic (QFLIA) is undecidable, even for transition systems and safety properties expressed in QFLIA.
\end{abstract}

\section{Introduction}

We address the problem of inferring inductive invariants in the language of quantifier free linear integer arithmetic (QFLIA).
Inductive invariants are the concept underlying most of the approaches for safety verification of infinite-state systems.
Inductive invariants in QFLIA are inferred by SMT-based tools such as Spacer~\cite{DBLP:conf/cav/KomuravelliGC14}. 

In this paper, we show that the problem of inferring such inductive invariants is undecidable even when the transition system at hand is expressed in the language of quantifier free linear integer arithmetic.

\section{The Problem of Inferring QFLIA Invariants}

We consider transition systems and properties expressed using \emph{quantifier free linear integer arithmetic} (QFLIA).
A transition system is represented by a tuple $\TS = (\Vars,\Init,\TR)$, where $\Vars$ is a set of integer variables, $\Init$ is an initial states formula  in QFLIA over $\Vars$, and $\TR$ is a transition relation formula in QFLIA over $\Vars \uplus \Vars'$ where $\Vars' = \{v' \mid v \in \Vars\}$ is a copy of the variables used to describe the target state of a transition. Each assignment to $\Vars$ defines a state of $\TS$, and each assignment that satisfies $\Init$ defines an initial state. The set of transitions is the set of all pairs of states $(s,t)$ satisfying $\TR$ (where the interpretation of the unprimed variables is taken from $s$ and the interpretation of the primed variables is taken from $t$). A safety property $\Prop$ of $\TS$ is expressed by a QFLIA formula over $\Vars$. We say that $\TS \models \Prop$ if all the reachable states of $\TS$ (defined in the usual way) satisfy $\Prop$.

An \emph{inductive invariant} $\Inv$ for $\TS = (\Vars,\Init,\TR)$ and $\Prop$ is a formula over $\Vars$ such that
\begin{inparaenum}[(i)]
\item $\Init \Rightarrow \Inv$ (initiation),
\item $\Inv \wedge \TR \Rightarrow \Inv'$ (consecution), and
\item $\Inv \Rightarrow \Prop$ (safety).
\end{inparaenum}
(where $\Inv'$ denotes the result of substituting $v$ with $v'$ in $\Inv$, for every $v \in \Vars$, and $\varphi \Rightarrow \psi$ denotes that $\varphi \to \psi$ is valid.)
It is well known that if there exists an inductive invariant $\Inv$ as above, then $\TS \models \Prop$, i.e., $\TS$ is safe.

In the sequel we are interested in the existence of inductive invariants expressed in QFLIA. We therefore consider the following decision problem:

\begin{definition}[Inference of QFLIA invariants]
The problem of \emph{inferring QFLIA invariants} is to determine whether a given transition system $\TS$ and a safety property $\Prop$, both expressed in QFLIA, have an inductive invariant $\Inv$ expressed in QFLIA.
\end{definition}

We show that this problem is undecidable.

\begin{remark}
Determining the \emph{safety} of a transition system expressed in QFLIA is trivially undecidable since a counter machine may be encoded via a transition system in QFLIA, and its nontermination is a safety property (expressible in QFLIA). That is, the complement of the halting problem of counter machines reduces to the safety problem of QFLIA transition systems in a straightforward manner.
However, the problem of inferring QFLIA inductive invariants is different from the problem of safety verification since a transition system (and in particular one that is expressible in QFLIA) may be safe but not have an inductive invariant in QFLIA (\Cref{fig:example} provides such an example as we discuss in \Cref{sec:warmup}).
Therefore, undecidability of the safety problem does not imply that the problem of inferring QFLIA inductive invariants is undecidable.
\end{remark}

\section{Undecidability of QFLIA Invariant Inference}

We prove that the problem of inferring QFLIA inductive invariants is undecidable by a reduction from the halting problem of $2$-counter machines.
We start with an example of a simple program (transition system) that has no inductive invariant in QFLIA.

\subsection{Warmup: Example of absence of inductive invariant in QFLIA.}
\label{sec:warmup}
Consider the program $\Prog$ depicted in \Cref{fig:example}.
We prove that $\Prog$ has no inductive invariant in QFLIA. We will later use $\Prog$ and a similar argument in the correctness proof of the reduction.

\begin{figure}
\footnotesize
\centering
\begin{lstlisting}
Prog(int x){
  assume (x > 0) ;
  int z1, z2, y1, y2;
  z1:=x; z2:=2*x; y1:=0; y2:=0;
  while (z1>0) {
    z1--; y1:=y1+x;
  }
  while (z2>0) {
    z2--; y2:=y2+x;
  }
  assert (y2 = 2*y1) ;
}
\end{lstlisting}
\caption{A program that computes $y_1 = x^2$ and $y_2 = 2x^2$ and asserts that $y_2 = 2y_1$. Verifying it requires a non-linear inductive invariant. \label{fig:example}}
\end{figure}

The program receives as input an integer $x$, computes $y_1 = x^2$ and $y_2 = 2x^2$ in two loops, and asserts that $y_2 = 2y_1$.
It is straightforward to encode $\Prog$ as a transition system $\TS_{\Prog}$ 
by adding an explicit program counter variable, denoted $\pc$. We consider the $\pc$ location $\pcstart$ prior to the first loop, the $\pc$ locations $\pcloopone$ and $\pclooptwo$ at the body of each of the loops, and the $\pc$ location $\pcend$ of the assertion.
We define the initial states formula to be $\Init := \pc = \pcstart \wedge x > 0 \wedge z_1 = x \wedge  z_2 =2x \wedge  y_1=0 \wedge  y_2=0$ (i.e., after the initialization of the variables) and the transition relation formula encodes all the transitions as expected (the granularity of transitions is defined by the above $\pc$ locations). Since all the expressions and conditions in the program are linear, both $\Init$ and the transition relation formula of $\TS_{\Prog}$ are in QFLIA.
The assert statement induces the safety property $\Prop := (\pc = \pcend) \to (y_2 = 2y_1)$, which is also in QFLIA.

It is easy to be convinced that for every possible initial value of $x$ (satisfying the precondition $x > 0$), the assertion holds. Therefore, $\TS_{\Prog} \models \Prop$.
However, as we prove next, there is no inductive invariant in QFLIA for $\TS_{\Prog}$ and $\Prop$, i.e., no inductive invariant in QFLIA is capable of verifying the program.

\begin{proof}
Consider the set $R$ of reachable states encountered when the first loop terminates.
This set consists of all states of the following form (we omit the value of the program counter), for $n >0$:
\begin{align*}
(x, z_1, z_2, y_1, y_2) \mapsto  (n, 0, 2n, n^2, 0) \\
\end{align*}

Recall that, due to the initiation and consecution properties, an inductive invariant must overapproximate the set of reachable states. Therefore, any inductive invariant formula must be satisfied by these states.
In order to show that no such formula exists in QFLIA, we show that any QFLIA formula that is satisfied by all of these states is also satisfied by a state that reaches a bad state in a finite number of steps (i.e., a state for which, after executing the second loop, $y_2 \neq 2y_1$). However, the safety and consecution properties imply that such a state must not exist (by induction on the length of the execution leading to a bad state). Therefore, we conclude that no QFLIA inductive invariant exists in this case.

By way of contradiction, let $\varphi = \varphi_1 \lor \ldots \lor \varphi_r$ be a QFLIA formula, written in disjunctive normal form (DNF), where each $\varphi_i$ is a cube (conjunction of literals), such that $\varphi$ is satisfied by all the states in $R$. Define $R_1,\ldots,R_r \subseteq R$ such that $R_i = \{s \in R \mid s \models \varphi_i\}$ includes all states in $R$ that satisfy $\varphi_i$. We show that there exists $i \in \{1,\ldots,r\}$ such that $\varphi_i$ (and hence $\varphi$) is satisfied by a state that reaches a bad state.

$R$ includes in particular all the states of the form $(n, 0, 2n, n^2, 0)$ where $n>0$ is an even number. Clearly, this is an infinite set of states. Therefore, since there are finitely many $R_i$'s that together cover $R$, there exists $i \in \{1,\ldots,r\}$ such that $R_i$ also includes infinitely many such states. We  view these states as vectors in a $5$-dimensional space. Take two such vectors
$(n, 0, 2n, n^2, 0)$ and
$(m, 0, 2m, m^2, 0)$  in $R_i$ where $n \neq m$. Then their linear combination
$(\frac{1}{2}(n+m), 0, n+m, \frac{1}{2}(n^2+m^2), 0)$ is in the convex hull of $R_i$. Therefore, it must satisfy $\varphi_i$ ($\varphi_i$ is a cube in QFLIA that is satisfied by all states in $R_i$, hence it is also satisfied by all states in its convex hull). Furthermore, the vector $(\frac{1}{2}(n+m), 0, n+m, \frac{1}{2}(n^2+m^2), 0)$ defines a valid state (all values are integers).

However, when executing the second while loop starting from the state $(x, z_1, z_2, y_1, y_2) \mapsto (\frac{1}{2}(n+m), 0, n+m, \frac{1}{2}(n^2+m^2), 0)$, the outcome is the state
$(x, z_1, z_2, y_1, y_2) \mapsto (\frac{1}{2}(n+m), 0, 0, \frac{1}{2}(n^2+m^2), (n+m)^2)$, and since $(n+m)^2 \neq n^2+m^2$ (recall that $n,m>0$), the resulting state violates the assertion.
Therefore, the state $(\frac{1}{2}(n+m), 0, n+m, \frac{1}{2}(n^2+m^2), 0)$ that satisfies $\varphi_i$ (and hence $\varphi$) reaches a bad state in a finite number of steps. This completes the proof.
\qed
\end{proof}

\subsection{The reduction}

We establish undecidability of inferring QFLIA invariants by a reduction from the halting problem of Minsky (2-counter) machines.
The general scheme of the reduction resembles the one used in~\cite{DBLP:conf/popl/PadonISKS16} to prove the undecidability of inferring universally quantified inductive invariants in uninterpreted first order logic for EPR transition systems.

The input of the reduction is an arbitrary Minsky machine, $M = (Q, c_1, c_2)$,
where $c_1$, $c_2$ are counters, both initially $0$, and $Q = \{q_1, \ldots, q_n\}$ is a
finite sequence of instructions, where $q_1$ is the first instruction, and
$q_n$ is the halting instruction. The possible instructions are:
\begin{itemize}
\item $i_k$: increment counter $c_k$,
\item $d_k$: decrement counter $c_k$,
\item $t_k(j)$: if counter $c_k$ is $0$ go to instruction $j$.
\end{itemize}
where in each case, control is passed to the next instruction except
when the tested counter is $0$ and thus the branch is taken.

The reduction constructs a transition system and safety property, both expressed in QFLIA, such that
the transition system has an inductive invariant in QFLIA if and only if $M$ halts.

To do so, the reduction constructs a program $\Progcomp$ that, on input $x >0$, runs $M$ (which ignores $x$) in parallel to running $\Prog$ from \Cref{fig:example} on $x$.
If $M$ terminates then $\Progcomp$ terminates.
If $\Prog$ terminates, $\Progcomp$ continues to run $M$ (unless $M$ has also terminated, in which case it terminates).
Clearly, the corresponding transition system is expressible in QFLIA over the variables $\pc, x, z_1, z_2, y_1, y_2,c_1,c_2,q$, where $\pc, x, z_1, z_2, y_1, y_2$ are the variables of $\TS_{\Prog}$ and $c_1,c_2,q$ are variables for the counters and control location of $M$. We denote the resulting transition system by $\TScomp$.
The safety property is inherited from $\Prog$, and is also expressible in QFLIA.
(In particular, if $M$ terminates before $\Prog$ terminates on $x$, then $\Progcomp$ terminates before reaching the assertion.)

While the product transition system $\TScomp$ is safe for every $M$, we show that if $M$ terminates, $\TScomp$ has an inductive invariant in QFLIA, whereas if $M$ does not terminate, it does not.

\begin{proposition}
\label{claim:red}
Let $M$ be a Minsky machine and $\TScomp$ be defined as above.
Then $\TScomp$ has a QFLIA inductive invariant if and only if $M$ terminates.
\end{proposition}

\paragraph{Proof.}
\paragraph{$\Leftarrow$:} Suppose $M$ terminates, say after $k$ steps.
Then the set of reachable states of $\TScomp$ can be expressed by a finite disjunction of ``cases'', as we explain next.

First, consider the reachable states that correspond to input values of $x$ that are smaller or equal than $k$. Since there are finitely many such values, and since we consider only a finite number of steps ($k$) on them, this set is finite. Each state in the set corresponds to the values of variables, counters and the control location, and can be characterized by a QFLIA cube. Hence the entire \emph{finite} set can be encoded precisely by a QFLIA formula $\varphi_{x \leq k}$ that comprises of a finite disjunction of the aforementioned cubes over all the reachable states. The set of states satisfying $\varphi_{x \leq k}$ is precisely the set of reachable states for $x \leq k$.

Now, consider the reachable states that correspond to input values of $x$ that are greater than $k$.
For such inputs, only the first loop in $\Prog$ will get to run (since $\TScomp$ will terminate after $k$ steps where $k$ is smaller than the value of $x$ which is also the number of iterations of the first loop),
and will produce the following reachable states, for every $n > k$ and number of steps $0< t \leq k$,
\begin{align*}
(\pc, x, z_1, z_2, y_1, y_2) \mapsto &\qquad (\pcloopone, n, n-t, 2n, nt, 0) \\
&\qquad (\pcstart, n, n, 2n, 0, 0) \\
\end{align*}
Each such state is augmented with the values of the counters and control location of $M$ after $t$ steps.
For every $0< t \leq k$, the states reachable in $t$ steps over inputs $x >k$ can therefore be characterized by the following QFLIA cube:
\[
\varphi_t := x > k \wedge \pc = \pcloopone \wedge z_1 = x-t \wedge z_2 = 2x \wedge y_1 = tx \wedge y_2 = 0 \wedge \varphi_{M(t)}
\]
and the states reachable in $0$ steps over inputs $x >k$ can be characterized by the following QFLIA cube:
\[
\varphi_0 := x > k \wedge \pc = \pcstart \wedge z_1 = x \wedge z_2 = 2x \wedge y_1 = 0 \wedge y_2 = 0 \wedge \varphi_{M(0)}
\]
where $\varphi_{M(t)}$ is the QFLIA cube representing $M$'s state after $t$ steps. (Note that $t$ is a fixed value hence the constraints are linear.)
Since $t$ is bounded by $k$, the entire set can be expressed by a finite disjunction, enumerating all possibilities of $t \leq k$:
\[
\varphi_{x>k} := \bigvee_{0 \leq t \leq k} \varphi_t
\]

Finally, the formula $\varphi_{x \leq k} \vee \varphi_{x>k}$ is a QFLIA inductive invariant for $\TScomp$. Consecution follows from the properties of the set of reachable states.

\paragraph{$\Rightarrow$:}
Suppose $M$ does not terminate. In this case, we use an argument similar to the one used in \Cref{sec:warmup} to show that $\Prog$ has no inductive invariant in QFLIA to show that the product system also does not.

As before, let $\varphi = \varphi_1 \lor \ldots \lor \varphi_r$ be a QFLIA formula, written in DNF form, where each $\varphi_i$ is a cube (conjunction of literals).
We show that if $\varphi$ is satisfied by all the reachable states of $\TScomp$, then there exists $i \in \{1,\ldots,r\}$ such that $\varphi_i$ (and hence $\varphi$) is satisfied by a state that reaches a bad state. As before, this implies that $\varphi$ is not an inductive invariant for $\TScomp$, and hence no inductive invariant in QFLIA exists.

For each $t >0$, consider the set $R_t$ of reachable states encountered after running $\TScomp$ exactly $t$ steps. The states in the set have the following form:
\begin{align*}
&(\pc, x, z_1, z_2, y_1, y_2,c_1,c_2,q) \mapsto\\
  & \qquad (\pcloopone,n, n-t, 2n, nt, 0, f_1(t), f_2(t), f_q(t)) &\mbox{ for $n \geq t$} \\
& \qquad (\pclooptwo,n, 0, 2n-(t-n), n^2, n(t-n), f_1(t), f_2(t), f_q(t))  &\mbox{ for $\frac{t}{3} < n< t$}\\
& \qquad (\pcend,n, 0, 0, n^2, 2n^2, f_1(t), f_2(t), f_q(t))  &\mbox{ for $0 < n \leq \frac{t}{3}$}\\
\end{align*}
where $f_1, f_2, f_q$ denote functions that map a number $t$ to the values of $c_1, c_2, q$, respectively, obtained after $t$ steps of $M$.

If $\varphi$ is satisfied by all the reachable states of $\TScomp$, then it is satisfied by all the states in $R_t$ (in particular).
Define $R^t_1,\ldots,R^t_r \subseteq R_t$ such that $R^t_i = \{s \in R_t \mid s \models \varphi_i\}$ includes all states in $R_t$ that satisfy $\varphi_i$.

Next, choose (arbitrarily) some $t$ such that $t \geq 3(r+2)$. Then the open interval 
$(\frac{t}{3}, t)$ contains $\geq 2(r+1)$ integer values, and in particular $\geq r+1$ even integer values.
Each of these values $n$ satisfies $\frac{t}{3} < n< t$,
and therefore corresponds to a state of the form $(\pclooptwo, n, 0, 2n-(t-n), n^2, n(t-n), f_1(t), f_2(t), f_q(t))$ in $R_t$.
Since there are more than $r$ such states but only $r$ sets $R^t_i$, by the pigeonhole principle, there exists $i \in \{1,\ldots,r\}$
such that there are at least two such states in $R^t_i$. We view these states as $9$-dimensional vectors. Say that these vectors are:
\begin{align*}
&\bar{v}_1 = \big(\pclooptwo, n,\ 0, 3n-t,\ n^2, \ nt-n^2,\ \  f_1(t), f_2(t), f_q(t)\big) \mbox{\quad and} \\
&\bar{v}_2 = \big(\pclooptwo, m, 0, 3m-t, m^2, mt-m^2, f_1(t), f_2(t), f_q(t)\big)
\end{align*}
where $n \neq m$, both $n,m$ are even and both of them are in the interval
$(\frac{t}{3}, t)$. Then their linear combination
\begin{equation} \label{combination-state}
\big(\pclooptwo, \frac{1}{2}(n+m), 0, \frac{3}{2}(n+m)-t, \frac{1}{2}(n^2+m^2), \frac{1}{2}(n+m)t-\frac{1}{2}(n^2+m^2), f_1(t), f_2(t), f_q(t)\big)
\end{equation}
is in the convex hull of $R^t_i$. Therefore, it must satisfy $\varphi_i$ ($\varphi_i$ is a cube in LIA that is satisfied by all states in $R^t_i$, hence it is also satisfied by all states in its convex hull).
Furthermore, the vector in \Cref{combination-state} defines a valid state (all values are integers since $n,m$ are even).

To conclude the proof, we show that when executing the program $\TScomp$ from the aforementioned state that satisfies $\varphi_i$ (and hence $\varphi$), a bad state is reached.
To this end, we first observe that the state of $M$ corresponds to a real reachable state $( f_1(t), f_2(t), f_q(t))$, hence, we are guaranteed that $M$ does not terminate when considering executions from this state (no matter how many steps are performed).
Now consider the execution of $\Prog$ starting from the state
\begin{align*}
&(\pc, x, z_1, z_2, y_1, y_2,c_1,c_2,q) \mapsto\\
 &\qquad (\pclooptwo,\frac{1}{2}(n+m), 0, \frac{3}{2}(n+m)-t, \frac{1}{2}(n^2+m^2), \frac{1}{2}(n+m)t-\frac{1}{2}(n^2+m^2))
\end{align*}
Since $n,m > \frac{t}{3}$, we get that $\frac{3}{2}(n+m)-t>0$. This is the value of $z_2$, the iterator of the second loop, hence the loop continues to iterate $\frac{3}{2}(n+m)-t$ iterations, where in each iteration it increments $y_2$ by $\frac{1}{2}(n+m)$ (the value of $x$), before it terminates and
the outcome is the state
\begin{align*}
(\pc,x, z_1, z_2, y_1, y_2) \mapsto (\pcend,\frac{1}{2}(n+m), 0, 0, \frac{1}{2}(n^2+m^2), v)
\end{align*}
where
\begin{align*}
v & =  \frac{1}{2}(n+m)t-\frac{1}{2}(n^2+m^2)+ \big(\frac{3}{2}(n+m)-t\big)\cdot\big(\frac{1}{2}(n+m)\big)\\
  & = \frac{1}{2}(n+m)t - \frac{1}{2}(n^2+m^2)+ \frac{3}{4}(n+m)^2 - \frac{1}{2}(n+m)t\\
  & = \frac{1}{4}(n^2+m^2) + \frac{3}{2}mn  
\end{align*}
is the value of $y_2$. To show that this is a bad state (violating the assertion $y_2 = 2y_1$), it suffices to show that
$v \neq n^2 + m^2$.
I.e., we require that $\frac{1}{4}(n^2+m^2)+ \frac{3}{2}mn  \neq n^2 + m^2$.
Simplifying the expressions, we see that this holds iff 
$(m-n)^2 \neq 0$, which is indeed the case since $m \neq n$.
This completes the proof.

\qed

\begin{remark}
Consider again the proof that non-termination of $M$ implies that there is no QFLIA inductive invariant for $\TScomp$.
In the proof we showed that if an inductive invariant $\Inv$ exists, then there are two certain points $\bar{v}_1$ and $\bar{v}_2$ whose linear combination defines a state that also satisfies $\Inv$ (by convexity) but also reaches a bad state, in contradiction to the properties of $\Inv$.
One may wonder why we did not select $\bar{v}_1$ and $\bar{v}_2$ as points encountered when the first loop terminates its execution (as is done in \Cref{sec:warmup}).
The problem is that these points would correspond to different states of $M$ (as they are encountered after different number of steps).  Hence, the linear combination might result in a state of $M$ that is not reachable and that $M$ terminates from. This would mean that $\TScomp$ may terminate its execution before reaching the assertion violation. Therefore, we cannot guarantee that the state defined by the linear combination reaches a bad state of $\TScomp$, and the proof fails.
\end{remark}

Since the termination problem of Minsky machines is undecidable, we conclude:

\begin{theorem}
The problem of inferring QFLIA invariants is undecidable (even for transition systems expressed in QFLIA).
\end{theorem}

\bibliographystyle{plain}
\bibliography{refs}

\end{document}